# A rotational traveling wave based levitation device
# – Modelling, design, and control

Ran Gabai [a,*], Dotan Ilssar [b], Ran Shaham [c], Nadav Cohen [d] and Izhak Bucher [e]

[a] Dynamics Laboratory, Faculty of Mechanical Engineering, Technion – Israel Institute of Technology, Haifa 3200003, Israel. rangab@technion.ac.il

[b] Dynamics Laboratory, Faculty of Mechanical Engineering, Technion – Israel Institute of Technology, Haifa 3200003, Israel. dilssar@technion.ac.il

[c] Dynamics Laboratory, Faculty of Mechanical Engineering, Technion – Israel Institute of Technology, Haifa 3200003, Israel. ransh@technion.ac.il

[d] Nadav Cohen – Engineering consulting and research Hatchelet 232, Haifa, Israel. nadav@minovation.org

[e] Dynamics Laboratory, Faculty of Mechanical Engineering, Technion – Israel Institute of Technology, Haifa 3200003, Israel. bucher@technion.ac.il

[*] Corresponding author:

Faculty of Mechanical Engineering, Technion City, Haifa 3200003, Israel

Tel: 972-77-8873605, email address: rangab@technion.ac.il

## ABSTRACT

Described is a device acting on an acoustically levitated object by manipulating the pressure and flow of a thin layer of air such that its rotation can be precisely controlled without mechanical contact. Virtual work analysis assists in simplifying the multi-actuator control problem into a problem governed by a controllable parameter. Actuation is done with a vibrating ring capable of producing ultrasonic standing and traveling waves, creating the acoustic excitation that affects the pressure in a thin, intermediate layer of gas. A distinctive vibration pattern is required to generate the temporal and spatial pressure field of the squeezed air layer that gives rise to both acoustic levitation force and rotational torque. Described are the physical and design development stages leading to an optimized structure, all followed by verifying and dynamics-calibration experiments. Moreover, by precisely controlling the ratio of standing and traveling waves in a closed-loop, one can affect the shear forces applied by the squeezed air layer, thus creating a non-contacting manipulation mechanism. An over-actuated set-up is converted via an algebraic transformation, into a simplified single control-parameter problem. The transformation ties the standing waves ratio, and hence the rotational torque, to the amplitudes and phases of the actuators. This arrangement leads to an effective closed loop methodology that was implemented experimentally showing good performance and exhibiting rapid angular positioning.

## 1 Introduction

Ultrasonic motors can make use of flexural propagating waves to induce tangential forces that generate propulsion opposite to the wave's travel (e.g. [1,2]). In near field acoustic levitation based motors [3,4], these forces are caused by viscous shear stresses of a thin air layer driven by pressure gradients inherent to the deformed traveling wave in the ring, affecting the air layer and thus the carried object. The present work deals with a unique type of ultrasonic motor, where vibrations of a stationary part (the stator) perform two tasks; (i) levitating the movable part (the rotor), forming a contactless bearing, and (ii) generating a rotating torque by propagating waves, controlling the motion of the rotor.

When two closely placed surfaces, surrounded by a compressible fluid, exhibit relative oscillatory normal motion, a layer of the ambient compressible fluid gets trapped in the thin clearance between the two surfaces. The formation of this thin layer, commonly known as "squeeze film", relies on the viscosity of the fluid, preventing the latter from flowing out of the clearance between the film's bounding surfaces. Moreover, due to the compressibility of the fluid, the pressure distribution inside the squeeze film fluctuates around a nominal profile whose average value is higher than the ambient pressure (e.g. [5–7]). Thus, the average force applied by the squeeze film is a repulsing force opposing gravity. As a result, a freely suspended planar object, representing the rotor, can be levitated above a rapidly oscillating surface, functioning as the stator. This phenomenon is often referred to as near-field acoustic levitation [3,8]. It should be noted that although the pressure inside the squeeze film is higher than the ambient pressure, the squeeze film maintains an average mass conservation in steady-state. Consequentially, the phenomenon described above can keep the rotor levitated at a steady-state around a mean height [9] of 5-200 micrometers with small rapid oscillation in the pico-meter range.

In addition to the normal vibrations discussed above, it was shown in earlier studies that traveling flexural waves of the rotor can be used to generate shear forces (e.g. [8,10,11]). Namely, flexural traveling waves of the stator cause pressure gradients at the wave's direction. These gradients result in flow along the waves' progression and therefore shear forces are generated, causing the rotor to perform the lateral motion. The squeeze film serves two purposes; as a contactless bearing – providing the levitation, and as a conveying element, propelling the rotor.

A special configuration is required in order to effectively excite progressing waves efficiently, as discussed, for example, in [4,12]. The device proposed here is based on a vibrating annulus creating the levitation and transportation forces efficiently. In [13], it is shown that in order to generate traveling flexural waves in an annulus, two actuators, phased in time and space, are required. The present paper proposes a different method to excite flexural traveling waves in an annulus shaped structure. By placing the actuators in an equally spaced distribution, the cyclic symmetry is maintained, reducing the modal contamination [14]; thus, near-pure traveling waves can be produced [15]. In order to tightly control the traveling waves, the mechanical design has to comply with several dynamical constraints, as well as the traveling wave control algorithm [16], and the real-time resonance tracking [17]. Successful operation is achieved by tightly integrating the dynamics and mechanical design with high-speed signal processing and control via a digital circuity (FPGA) driving the piezoelectric actuators through the high-voltage amplifiers.

This paper describes a motor, built and operated according to physical consideration while employing the abovementioned disciplines. The paper starts with a description of the motor's general structure and working principle. Next, a dynamic model of the system is presented and analyzed in order to develop the proper excitation and control. Finally, a series of experiments, proving the concept and verifying the dynamic model are provided.

## 2 Motor Structure, Modelling, and Design

The device dealt with in this paper, consists of three core components, a stationary part producing the required excitation, a freely suspended rotor, and a thin air layer squeezed between them. The general structure of the stator, designed to generate and maintain a steady squeeze-film and the required vibrating waves for propulsion, is displayed in Figure 1. As seen from the figure, the stator is composed of an array of three vibrating actuators (Langevin type transducer model FBL28452HS with the natural frequency of 28kHz), connected along the circumference of an aluminum annulus (inner diameter 100mm, outer diameter 150mm, thickness 5mm). The rotor is a freely suspended disk placed above the vibrating annulus. Additional details regarding the system are provided later in section 4.1. In the experiment, it has an axial bearing to maintain its alignment with the annulus while rotating, and an encoder is measuring its angular movement relative to the annulus. High frequency (ultrasonic) vibrations of the annulus generate an acoustic field in the air gap, levitating the rotor above the annulus. Furthermore, by controlling the amplitudes and phases of the actuators, tangential traveling pressure waves are formed to induce the rotation of the rotor. It should be noted that in the intended application, nor a mechanical bearing or any mechanical contact will be necessary and the levitated object will be held only by air pressure. For this purpose, several such rings will operate in unison while being controlled independently [18].

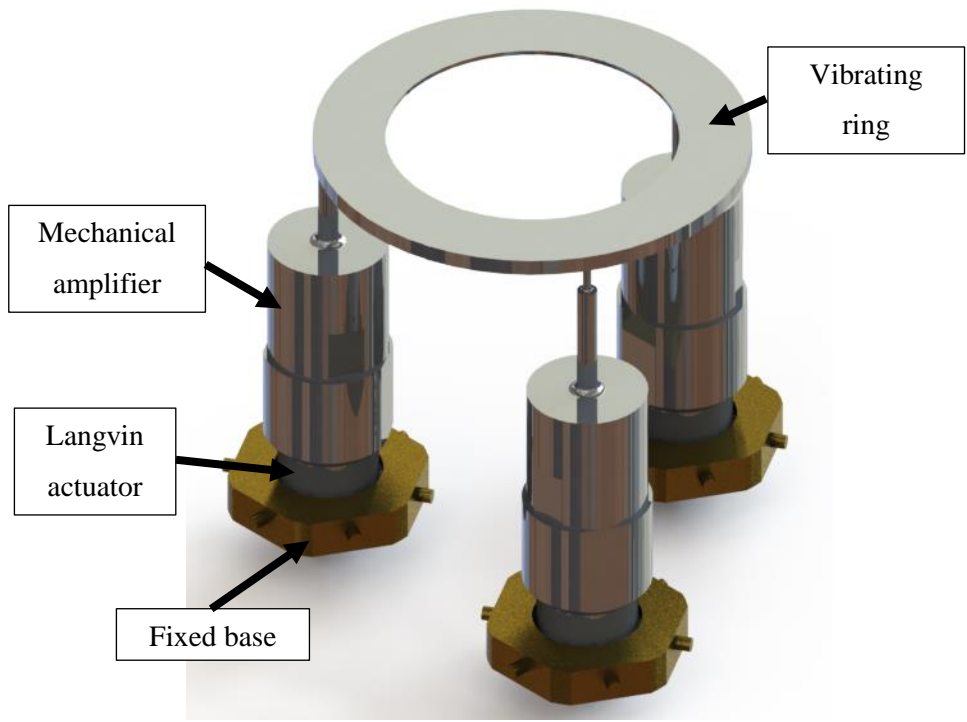

Figure 1. The general structure of the stationary part of the motor (stator). The three actuators are designed to resonate at the same frequency as the Langevin actuator consisting of their bottom part.

## 2.1 Annulus design

In order to generate both levitation and propulsion, the desired operating shape of the stator should exhibit bending vibrations in the tangential direction. Due to the cylindrical symmetry, the annulus' modes of vibrations (or eigenvectors), with at least one nodal diameter, appear in pairs (doublets) for every relevant natural frequency [19]. The doublet modes have identical spatial response amplitudes and wavelengths but are phased by $\pi/2$ in space. In general, the $k^{th}$ pair of tangential doublets modes can be written as:

$$\phi_{c,k}(r,\theta) = A_c \cos(k\theta)R(r), \qquad \phi_{s,k}(r,\theta) = A_s \sin(k\theta)R(r), \qquad (1)$$

$\phi_{c,k}, \phi_{s,k}$ are referred as the cosine and sine modes. The term $R(r)$ represents a combination of Bessel functions, describing deflection in the radial deflection of the annulus [19]. Figure 2 shows several mode shapes near the operating frequency of a free, uniform thickness annulus, calculated with a finite elements software. The radial effect described by $R(r)$ is clearly seen for any of the presented mode-shapes.

When the annulus is excited at a frequency near a natural frequency, the corresponding two modes, related to the $k^{th}$ pair of modes, are nearly the only ones being dynamically excited. Indeed, any combination of these modes can form standing and traveling waves [20], all at high vibration amplitudes. Clearly, being able to excite the structure at resonance, increases the efficiency of the required force and enhances the ability to generate a squeeze film with sufficient pressure elevation. Moreover, for the sake of high efficiency, the natural frequency of the annulus is matched to the actuator's natural frequency to assure impedance matching.

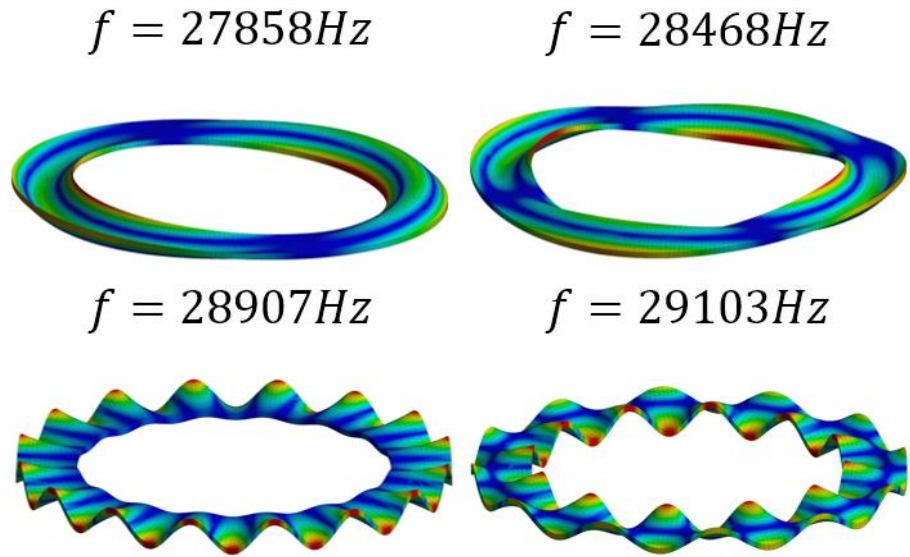

Figure 2. Finite elements based modes of a uniform annulus in the vicinity of the operating frequency

Further optimization of the squeeze-film generating surface is achieved by tapering the cross section to render the response uniform in the radial direction, obtaining $R(r) \approx \text{const}$.

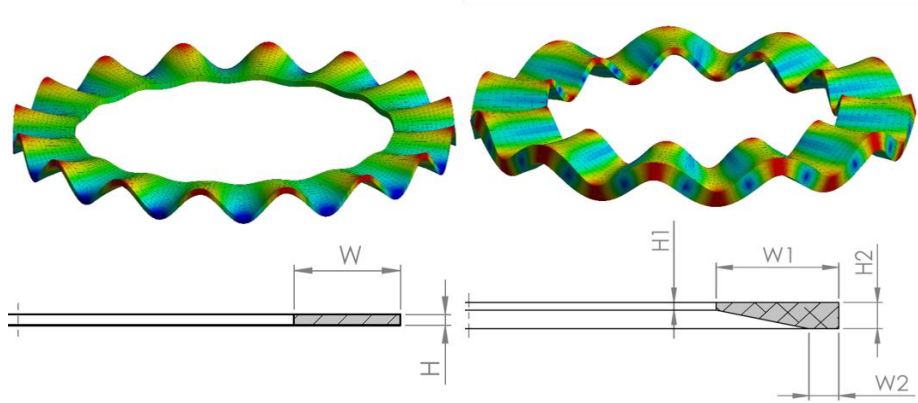

Figure 3. Finite elements results of the response of tapered and untampered annulus.

Figure 3 provides a graphical comparison of the annulus mode for an un-tapered $(W = 25mm, H = 5mm)$ and tapered annulus $(W_1 = 25mm, W_2 = 10mm, H1 = 2.5mm, H2 = 5mm)$.

## 2.2 Excitation of the annulus

In general, attaching an actuator to the annulus breaks the cylindrical symmetry. Clearly, the latter can cause the doublet modes to split [21], meaning that the natural frequencies corresponding to the doublet modes of the new structure will not be equal. Another phenomenon that can arise due to the symmetry-breaking is the spatial contamination of the doublet mode shapes with additional wavelengths [14,22]. Therefore, it is desired to locate the actuators in a way that maintains some of the symmetry related behavior. This can be achieved by equally spacing the actuators along the circumference of the annulus in a particular way. For *n* actuators, the angular spacing between every two actuators would be $2\pi/n$.

In order to excite a traveling wave, both the doublet modes have to be excited, implying that at least two actuators are necessary. Due to the orthogonality of the sine and cosine modes, it is not possible to excite both of them with two actuators spaced at an angle of $\pi$. This means that at least three actuators are required for maintaining cyclic symmetry and for being able to excite traveling waves (that are a combination of the sine and cosine modes phased at 90 degrees apart in time [23]). Indeed, as seen in Figure 1, the device discussed in the present paper consists the minimal number of actuators, spaced by an angle of $2\pi/3$. Moreover, for the sake of high efficiency, the impedance of the actuators was matched to the impedance of the annulus in a way that the natural frequencies of the actuators and the annulus are identical. The impedance matching was done by altering the dimensions of the mechanical parts while repeating the finite element analysis until convergence.

### 2.2.1 Exciting the annulus in a desired pattern of vibrations

In general, the forced response of the annulus can be computed as the sum of all its modes, each mode with its own temporal response. The response can be written as:

$$w(\theta,t) = \eta_0(t)\phi_o + \sum_{n=1}^{\infty} \eta_{c,n}(t)\phi_{c,n}(\theta) + \sum_{n=1}^{\infty} \eta_{s,n}(t)\phi_{s,n}(\theta), \tag{2}$$

where the modal coordinates $\eta_0, \eta_{c,n}, \eta_{s,n}$ depend on the excitation, and the term $\phi_0$ stands for the rigid body mode of the annulus. When the annulus is excited near the natural frequency, $\omega_k$, corresponding to the $k^{th}$ pair of doublet modes, all participation factors of the other modes become negligible, thus Eq.(2) becomes:

$$w(\theta,t) = \eta_c(t)\phi_c(k\theta) + \eta_s(t)\phi_s(k\theta) + \eta_0(t)\phi_0. \tag{3}$$

The desired vibration pattern of the annulus is achieved by controlling the phase and amplitude of the three actuators, represented as three forces given by:

$$f_r(t) = F_r \cos(\omega_k t - \varphi_r), \quad r = 1,2,3, \tag{4}$$

where $F_r, \varphi_r$ are the amplitude and time shift of the $r^{th}$ actuator, respectively. The total force applied by the actuators on the annulus is expressed as:

$$F(\theta,t) = \sum_{r=1}^{3} f_r(t)\delta\left(\theta - \frac{2\pi(r-1)}{3}\right), \tag{5}$$

where $\delta(\bullet)$ is the Dirac delta function of $\bullet$. As can be seen from Eq.(5) the spatial location of the first actuator is selected as a reference such that $\varphi_1 = 0$.

The principle of virtual work is applied to calculate the generalized forces, $Q_r$, corresponding to each mode of vibration, and a virtual displacement in the relevant modal coordinate, $\delta\eta_r$, as required to create the desired annulus response:

$$\delta W = F(\theta,t)\delta w = \sum_{r=1}^{3} Q_r \delta\eta_r(t). \tag{6}$$

The virtual displacement of the annulus can be calculated from Eq.(3):

$$\delta w = \delta\eta_1 A_c \cos(k\theta) + \delta\eta_2(t) A_s \sin(k\theta) + A_0 \delta\eta_3(t), \tag{7}$$

where $A_0$ is the response related to rigid body or other types of uniform motion. Calculating the virtual work from Eqns. (5), (6), (7):

$$\delta W = \sum_{r=1}^{3} \int_0^{2\pi} f_r(t)\delta\left(\theta - \frac{2\pi(r-1)}{3}\right)(\delta\eta_1 A_c \cos k\theta + \delta\eta_2(t) A_s \sin k\theta + A_0 \delta\eta_3(t))\mathrm{d}\theta. \tag{8}$$

Evaluating Eq.(8) yields:

$$\delta W = \sum_{r=1}^{3} f_r(t)\left(\delta\eta_1(t) A_c \cos\left(\frac{2\pi k(r-1)}{3}\right) + \delta\eta_2(t) A_s \sin\left(\frac{2\pi k(r-1)}{3}\right) + A_0 \delta\eta_3(t)\right). \tag{9}$$

Collecting the virtual displacement of each generalized coordinate, and equating to the virtual work of the generalized forces $(\delta\eta_c Q_c + \delta\eta_s Q_s + \delta\eta_0 Q_0)$, results in a transformation **T** relating the physical forces at the actuators and the generalized, mode related forces:

$$\begin{bmatrix} Q_c \\ Q_s \\ Q_0 \end{bmatrix} = \underbrace{\begin{bmatrix} A_c & A_c \cos\left(\dfrac{2\pi k}{3}\right) & A_c \cos\left(\dfrac{4\pi k}{3}\right) \\ 0 & A_s \sin\left(\dfrac{2\pi k}{3}\right) & A_s \sin\left(\dfrac{4\pi k}{3}\right) \\ A_0 & A_0 & A_0 \end{bmatrix}}_{\mathbf{T}} \begin{bmatrix} f_1 \\ f_2 \\ f_3 \end{bmatrix}. \tag{10}$$

The modes' coefficients $A_c, A_s, A_0$ can be scaled by setting:

$$A_c = A_s = \sqrt{\dfrac{2}{3}}, \quad A_0 = \sqrt{\dfrac{1}{3}}. \tag{11}$$

Consequently, the matrix $\mathbf{T}$ becomes unitary, meaning that $\mathbf{T}^{-1} = \mathbf{T}^T$.

Inverting Eq.(10) allows calculating the required forces applied by the actuators in order to generate any desired response on the annulus while working in the designated natural frequency:

$$\begin{bmatrix} f_1(t) \\ f_2(t) \\ f_3(t) \end{bmatrix} = \mathbf{T}^{-1} \begin{bmatrix} Q_c \\ Q_s \\ Q_0 \end{bmatrix}. \tag{12}$$

The generalized forces $Q_c, Q_s$ determine the annulus's vibration patterns and $Q_0$ is forced to zero in order to cancel the effect of rigid body modes. It should be mentioned the present approach can easily be extended to $n > 3$ actuators. This transformation reduces the evidently over-actuated problem with three actuators into a control problem with two variables. These variables generate the three required forces through Eq.(12). Observing the transformation $\mathbf{T}$, one may note that it is analogous to the Clarke & Park transformations [24], utilized when controlling a three-phase motor. The three actuators are the equivalent of the three motor poles and the combined doublet modes represent the rotating magnetic field.

### 2.2.2 Waves identification

During the calibration of the system, it is required to evaluate and quantify the traveling waves vibrating in the annulus. Considering the annulus vibration from Eq.(3), it was shown in [15,16], that the generalized coordinates $\eta_c(t), \eta_s(t)$ reside on an ellipse in the phase plane. While $\eta_0$ is an offset of the origins of the ellipse. Finding the ellipse parameters by curve-fitting, the ratio of traveling and standing vibrating waves can be obtained, even in real time. The ellipse parametric equations on the phase plane are:

$$\begin{aligned} \eta_1(t) &= a\cos(\omega t) + a_0 \\ \eta_2(t) &= b\sin(\omega t) + b_0, \end{aligned} \tag{13}$$

where $a, b$ are the radii of the ellipse and $a_0, b_0$ indicates the offsets from the origin. A common way to describe how far the actual response from a pure standing wave is by calculating a scalar measure called Standing Wave Ratio (*SWR*) [15]. A pure traveling wave yields $SWR = \pm 1$ where the sign indicates the direction of propagation while a pure standing wave yields $SWR = \infty$.

It is sometimes more convenient to use the inverse of this ratio since it is bounded between $-1 \leq SWR^{-1} \leq 1$ with $SWR^{-1} = 0$ corresponding to a pure standing wave. A different approach, quantifying the quality of the wave is the Traveling Wave Ratio (*TWR*), which can be calculated from the ellipse principle axes by [25]:

$$TWR = \frac{|a-b|}{a+b}, \qquad (14)$$

when $TWR = 0$ the system oscillates as a pure traveling wave and while $TWR = 1$, takes place for a pure standing wave. There is a simple relation between the *SWR* and *TWR*: $|SWR^{-1}| = 1 - TWR$.

### 2.2.3 Exciting waves in the annulus

In order to excite the annulus at a specific *SWR*, the doublet modes are excited such that their combination creates the desired vibration wave. This section presents the expressions of the forces required for exciting the annulus as standing and progressive flexural wave.

The annulus's response given by Eq.(3) describes a propagating wave when it takes the form:

$$w(\theta, t) = W_0 \left( \cos(\omega_k t) \cos(k\theta) \mp \sin(\omega_k t) \sin(k\theta) \right) = W_0 \cos(k\theta \pm \omega_k t), \qquad (15)$$

where the sign indicates the direction of propagation. Generating such a wave means setting the generalized forces such that:

$$Q_c = B_0 \cos(\omega_k t), \quad Q_s = \pm B_0 \sin(\omega_k t). \qquad (16)$$

By substituting Eq.(16) into Eq.(12), the forces for the three actuators are obtained:

$$\begin{bmatrix} f_1 \\ f_2 \\ f_3 \end{bmatrix} = A_c B_0 \begin{bmatrix} 1 \\ \cos\left(\frac{2\pi k}{3}\right) \\ \cos\left(\frac{4\pi k}{3}\right) \end{bmatrix} \cos(\omega_k t) \pm A_s B_0 \begin{bmatrix} 0 \\ \sin\left(\frac{2\pi k}{3}\right) \\ \sin\left(\frac{4\pi k}{3}\right) \end{bmatrix} \sin(\omega_k t). \qquad (17)$$

Since $A_c = A_s$ (see Eq.(11)), one can write $\tilde{B} = A_c B_0 = A_s B_0$. Further expanding Eq.(17) yields:

$$\begin{bmatrix} f_1 \\ f_2 \\ f_3 \end{bmatrix} = \tilde{B} \begin{bmatrix} \cos(\omega_k t) \\ \cos\left(\frac{2\pi k}{3}\right)\cos(\omega_k t) \pm \sin\left(\frac{2\pi k}{3}\right)\sin(\omega_k t) \\ \cos\left(\frac{4\pi k}{3}\right)\cos(\omega_k t) \pm \sin\left(\frac{4\pi k}{3}\right)\sin(\omega_k t) \end{bmatrix} = \tilde{B} \begin{bmatrix} \cos(\omega_k t) \\ \cos\left(\omega_k t \mp \frac{2\pi k}{3}\right) \\ \cos\left(\omega_k t \mp \frac{4\pi k}{3}\right) \end{bmatrix}. \qquad (18)$$

The forces obtained in Eq.(18) show that generating a propagating wave in the positive $\theta$ direction requires setting the forces induced by the actuators to have identical amplitude $\left( F_1 = F_2 = F_3 = \tilde{B}/A_c \right)$ and their phases should be shifted in time by $2\pi/3$ $(\varphi_1 = 0, \varphi_2 = 2\pi/3, \varphi_3 = 4\pi/3)$. To invert the direction of propagation, the phases of the 2nd and 3rd actuators are switched $(\varphi_1 = 0, \varphi_2 = 4\pi/3, \varphi_3 = 2\pi/3)$. Utilizing Eq.(12), the forces $f_1, f_2, f_3$ can be calculated in order to excite any excitation form, composed from the selected doublet mode of the annulus.

## 2.3 Forces generated by the squeeze-film

The behavior of the compressible air layer residing between the stator and the rotor is commonly approximated using Reynolds equation (e.g [6]). Here, the stator does not exhibit lateral motions, and the rotational velocity of the rotor is much slower than the traveling waves generated on the stator, implying that the former is much slower than the velocity of the entrapped gas. Consequentially, the lateral motions of the bounding surfaces can be disregarded when approximating the behavior of the gas. Under the abovementioned assumption, and under the assumption of isothermal conditions, Reynolds equation takes the following form [26]:

$$\frac{\partial}{\partial r}\left(rh^3 p \frac{\partial p}{\partial r}\right) + \frac{1}{r}\frac{\partial}{\partial \theta}\left(h^3 p \frac{\partial p}{\partial \theta}\right) = 12\mu r \frac{\partial (ph)}{\partial t}, \qquad (19)$$

where $t$ denotes the time, $r, \theta$ are the radial and tangential coordinates, respectively, and $p(r,\theta,t)$ symbolizes the pressure distribution inside the gas layer, where the layer's thickness is denoted $h(r,\theta,t)$, and $\mu$ stands for the dynamic viscosity of the fluid.

The squeeze film generates both the levitation force and the rotating torque. The normal force, acting in the axial direction in order to levitate the rotor. This force is obtained by integration of the gauge pressure of the entrapped gas, over the lateral area of the gas layer:

$$F_L = \int_0^{2\pi} \int_{r_{in}}^{r_{out}} r(p - p_a) \, dr d\theta. \qquad (20)$$

The torque is calculated by integrating the shear stresses exerted on the rotor surface area by the gas layer. The torque is expressed as a function of $h, p$, as suggested in [27], yields the following expression:

$$M \approx -\int_0^{2\pi}\int_{r_{in}}^{r_{out}} \frac{hr}{2}\frac{\partial p}{\partial \theta} \, dr d\theta. \qquad (21)$$

This expression allows calculating the rotating moment applied to the rotor. A simulation of the system was carried out, under a typical constant levitation gap and excitation frequency, where Eq.(21) was used in order to calculate the torque applied to the rotor. Figure 4 shows the torque obtained under different standing wave ratios at the stator, showing the monotonic change with the change of the quality of the traveling wave (*TWR/SWR*), excited within the annulus.

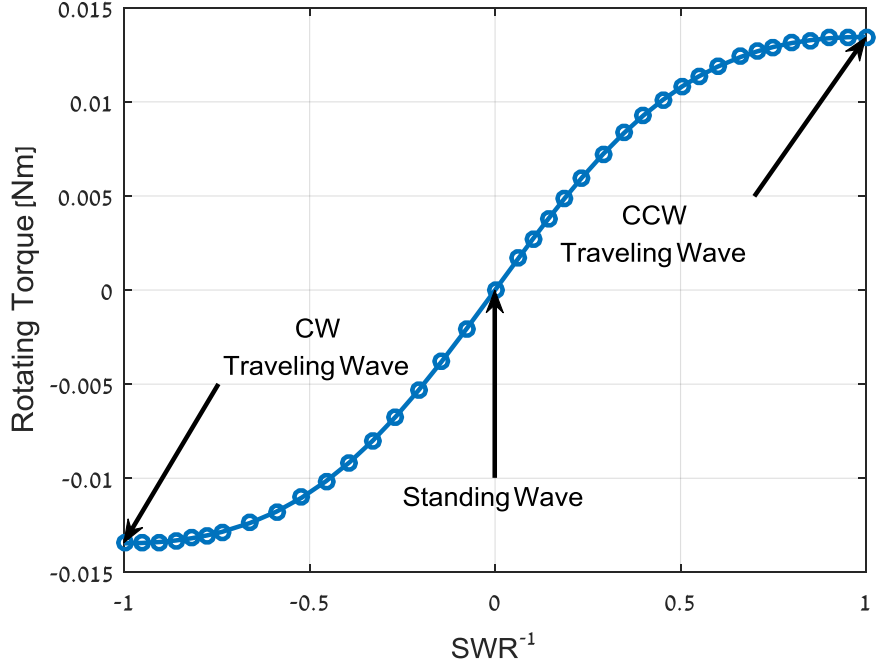

Figure 4. The numerically simulated rotating torque applied on the rotor, under excitation with different standing wave ratios.

Figure 4 was generated by approximating Eq.(19) via a finite difference approximation in space and adaptive integration in time in a similar manner to [28].

### 2.4 Rotor model – a rotating levitated disk

The rotor in this paper consists of a disk having mass $m$ and moment of inertia $J$, respectively. On the vertical axis, the rotor is subjected to the gravity force, a height-dependent levitation force denoted $F_L$ (see [7]), and damping forces. Thus, the equation of motion governing the levitation gap, $h$, is:

$$m\ddot{h} + c_h \dot{h} + mgh = F_L(t), \quad (22)$$

where $c_h$ is the damping coefficient, and $g$ is the gravitational acceleration. On the tangential axis, the disk experiences rotating torque, $M$, rotating the disk around its axis in an angle, denoted $\theta$. The angular equation of motion is:

$$J\ddot{\theta} + F_d(\dot{\theta}) = M(t), \quad (23)$$

where $M(t)$ is the rotating moment applied from the stator and $F_d$ represents dissipation torques. These torques are relative to the angular velocity by [29]:

$$F_d = c_2 \operatorname{sign}(\dot{\theta})\dot{\theta}^2 + c_1\dot{\theta}, \quad (24)$$

where $c_1$ is the coefficient of the linear friction and $c_2$ is the coefficient of drag forces. The $\operatorname{sign}(\ )$ operator aligns the direction of drag forces with the direction of rotation.

# 3 Angular positioning of the rotor

Controlling the position of the rotor involves regulating the vibrations of the annulus, measuring the annulus vibrations and, measuring the position of the rotor. The source of the rotating moment applied to the rotor is the propagating waves' vibrations of the annulus. Thus, it is necessary to tune the wave propagation in order to regulate the rotating moment. Tuning the waves allows controlling the direction and magnitude of the rotating moment. This section describes the tuning process of the actuation, in order to excite the annulus at the desired response. This paves the way toward closed loop control.

## 3.1 Exciting waves vibrations in the stator

Locating sensors above each actuator, spatially phased by $2\pi/3$ from each other, the annulus response to the excitation of the actuator is measured. Exciting all actuators in a single frequency and utilizing Eq.(3), the sensors' reading can be written as:

$$s_i = \eta_c(t)\phi_c(\theta_i) + \eta_s(t)\phi_s(\theta_i) + \eta_0(t)\phi_0, \quad i=1,2,3$$
$$\theta_i = (i-1)2\pi/3. \tag{25}$$

Reordering Eq.(25) into a matrix form, it can be seen that the transformation between the sensors' reading to the modal coordinates is identical to the transformation between the actuators' forces to the generalized forces **T** obtained at Eq.(10). Applying this transformation to the sensors' output, the modal coordinates can be extracted:

$$\begin{bmatrix} \eta_c \\ \eta_s \\ \eta_0 \end{bmatrix} = \mathbf{T} \begin{bmatrix} s_1 \\ s_2 \\ s_3 \end{bmatrix}. \tag{26}$$

By setting the excitation amplitudes and phases of the actuators forces (Eq.(4)), and utilizing the transformation **T** (Eq.(10)), the generalized forces are calculated. Applying these forces allows calculating the modal coordinates, $\eta_c, \eta_s$. By curve-fitting an ellipse to the measured response, the TWR can be calculated for any set of excitation forces. The *SWR/TWR* are a function of the relative amplitudes and phases between the actuators and can be controlled by setting the relative phase between the three actuators.

Assuming all actuators are identical (having the same magnitude), selecting $\varphi_1 = 0$ as a reference phase and modifying the two other phases $(\varphi_2, \varphi_3)$, a map of the *TWR* can be produced. A simulated map for $k=11$ is shown in Figure 5. The map is symmetric along a line from $\varphi_2 = 0, \varphi_3 = 0$ to $\varphi_2 = 2\pi, \varphi_3 = 2\pi$. Along this line, the value of *TWR* is one. This mean that the annulus is vibrating in a standing wave. Below this line the wave propagates in a clockwise direction and above it to the opposite direction.

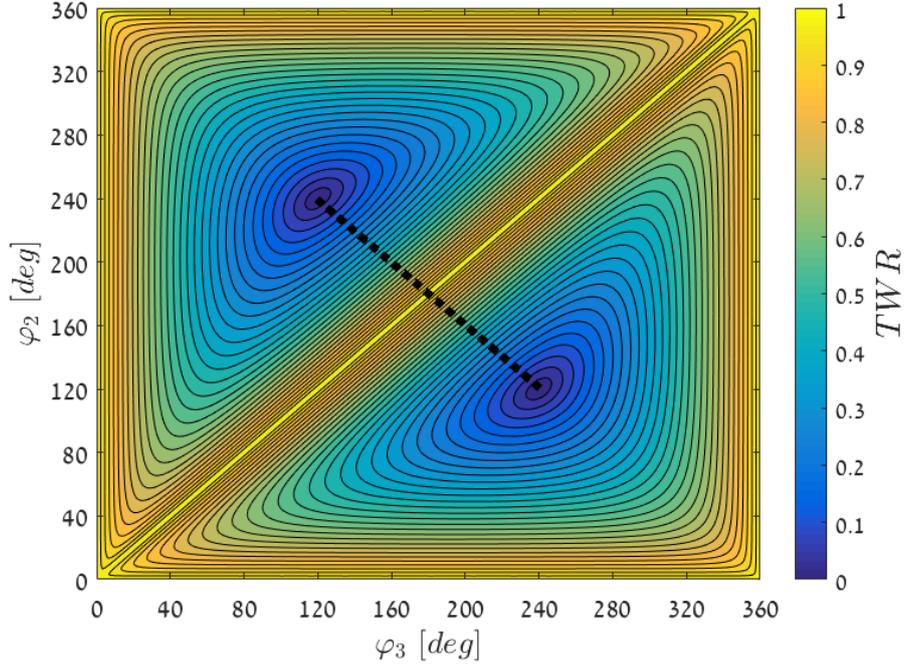

Figure 5. Simulated map of the *TWR* of an annulus (k=11) vs. the actuators phases. The dashed line is the wave control path.

The map has two minima points $(TWR = 0)$ located at $\varphi_2 = 2\pi/3, \varphi_3 = 4\pi/3$ and at $\varphi_2 = 4\pi/3, \varphi_3 = 2\pi/3$. Each minimum indicates the conditions to excite a traveling wave in a different direction (either clockwise or counter-clockwise). The latter complies with the theoretical results developed in section 2.2.3 (Eq.(18)).

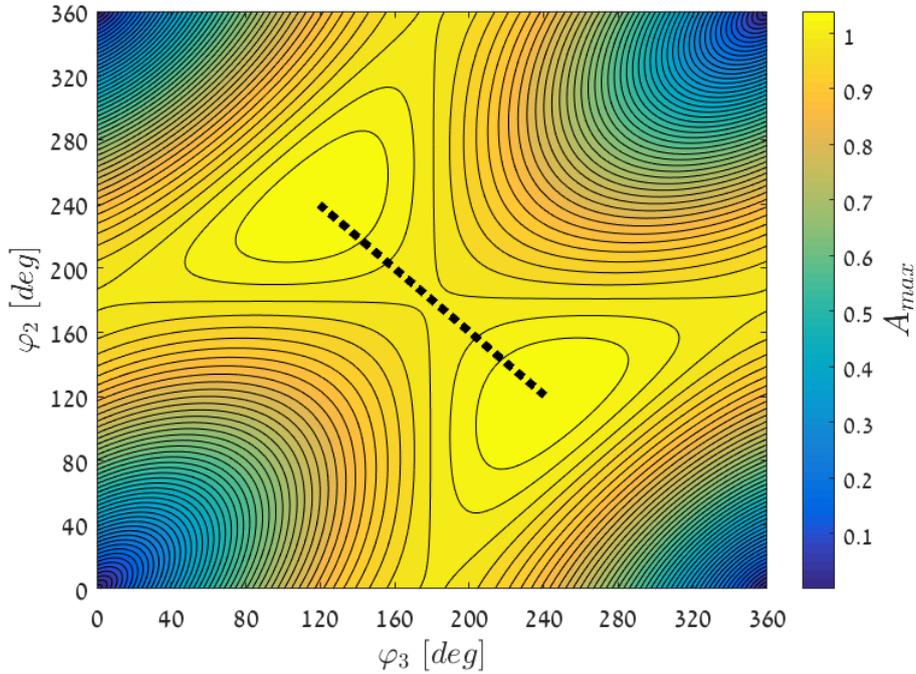

Figure 6. Simulated wave amplitude for *k=11* vs. the actuators phases. The dashed line is the wave control path.

An additional map, displays the normalized maximal amplitude of vibration, for the different actuator's phases is presented in Figure 6. The amplitude map provides additional significant information since the levitation forces are related to the amplitude of vibrations. Under low amplitudes, these forces might not be enough to support the levitation. As can be seen from Figure 5-Figure 6, it is advantageous to move along the dashed line in the $\varphi_2, \varphi_3$ domain. The linear relationship between $\varphi_2$ and $\varphi_3$ further reduces the number of the control parameters to one, as discussed below.

*3.2 Wave control*

Regulation of the propulsion forces is carried out by modifying the phases of the actuators to tune the *TWR* to the desired value. In addition, since the levitation height changes the behavior of the system, it should be kept approximately constant. By combining the information from both the *TWR* and amplitude maps, the optimal path is selected. This path is marked as a dashed line over the maps given at Figure 5-Figure 6. Choosing $u \in [-1,1]$ as the control parameter, the selected path is given by:

$$\varphi_2(u) = \pi/3 \cdot u + \pi, \quad \varphi_3(u) = -\pi/3 \cdot u + \pi. \tag{27}$$

The *TWR* vs. *u* for the selected path is shown in Figure 7(a). Setting $u = \pm 1$ sets the excitation phases for a pure traveling wave in either clockwise or counter-clockwise direction, whereas setting $u = 0$ provides the conditions for a pure standing wave.

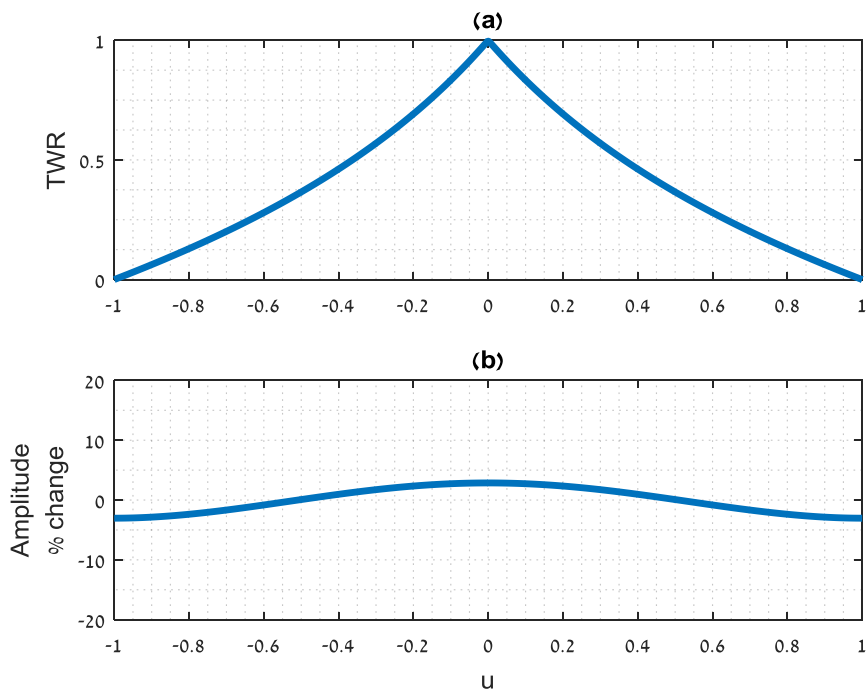

Figure 7. (a) Traveling Wave Ratio vs. u for the selected control path. (b) The change in amplitude in percentage for the selected path.

One can notice that along this path, a minimal change in the amplitude occurs. Figure 7(b) shows the amplitude change vs. *u*. It is seen that the deviation in the amplitude is no more the 3%.

In reality, differences in the actuators performance may detune the amplitudes of the forces, applied to the annulus. This affects the generation of the waves, hence, the *TWR* map is detuned as well. Figure 8 presents the *TWR* and maximum amplitude maps for the case where $|f_2|=0.8|f_1|$, $|f_3|=0.6|f_1|$. The *TWR* map (shown in Figure 8) is twisted in a way that the optimal phases between the actuators, which generates a traveling wave, deviate from the theory. In addition, inspecting the *TWR* values, they do not reach pure traveling wave $(TWR=0)$ at the minimum points.

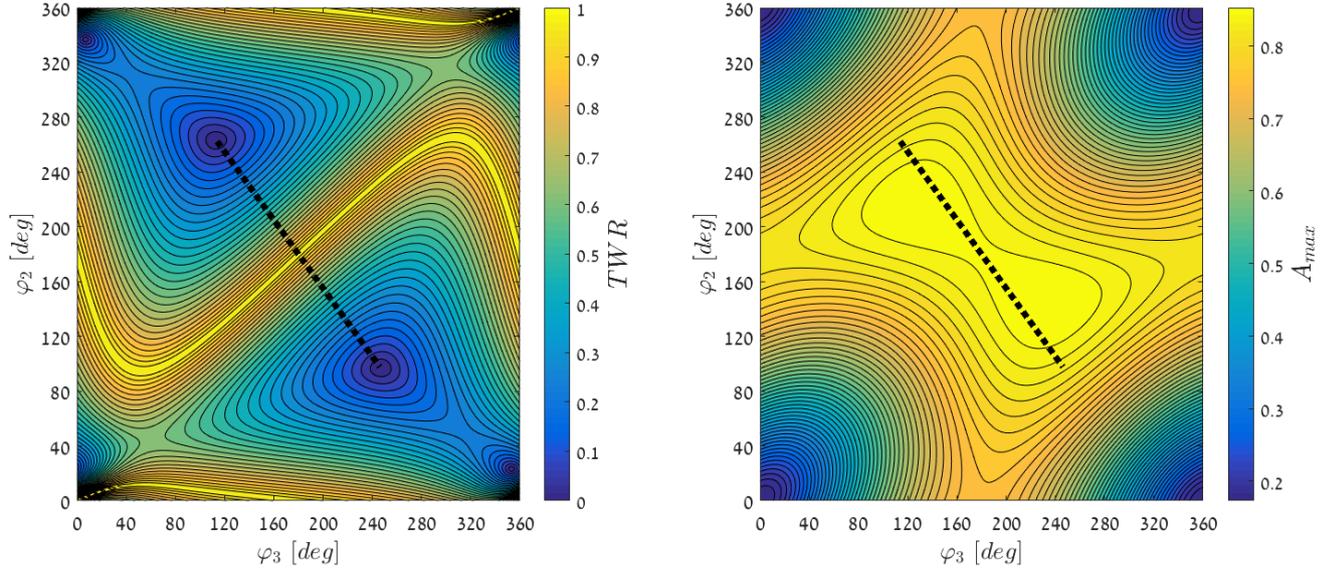

Figure 8. *TWR* (left) and maximum amplitude (right) maps for detuned forces amplitudes. The dashed line is the wave control path.

The difference in the forces amplitude prevents us from achieving a pure traveling wave for such setting as can be seen from Figure 9(a). The amplitude variation vs. the controlled parameter is shown in Figure 9(b). Comparing this to Figure 7(b), one can observe that the deviation of the amplitude of vibrations is within the same boundaries. In order to achieve pure traveling waves, the amplitude of the forces must be balanced. Additional sources that may detune the *TWR* map could be manufacturing tolerances of the annulus and horns that will break the cyclic-symmetric structure. The detuned effect of the *TWR* map requires the calibration of each stator individually in order to obtain the control path.

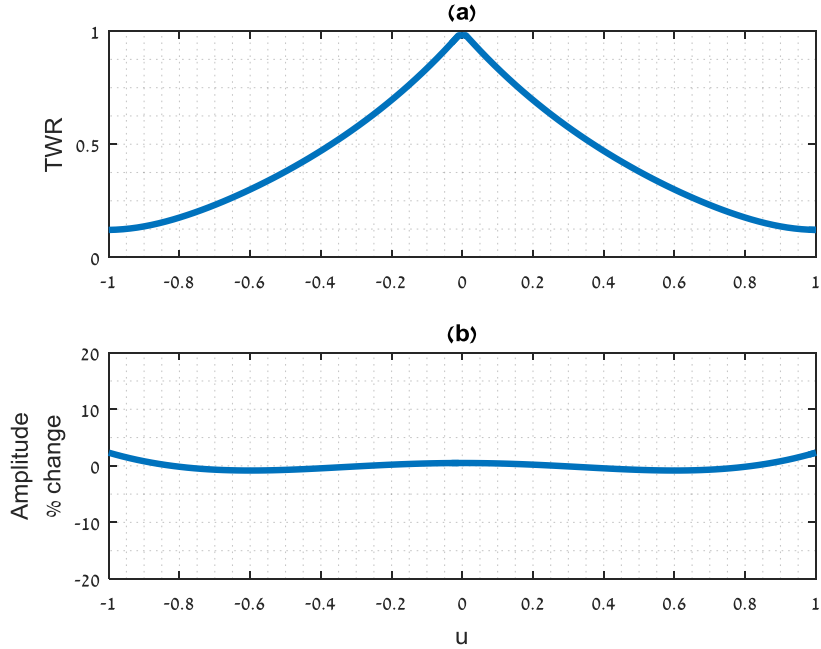

Figure 9. *TWR* (a) and maximum amplitude values (b) along the control line *u* for detuned forces amplitude

# 4 Experimental verification

The goal of the experiments that are described and their results is to prove the concepts which the device described in this paper is based upon. The principle of generating levitation and rotation by controlling the traveling waves in the annulus was demonstrated and compared to numerical results. In addition, closed-loop control was applied to demonstrate working in a closed loop.

## *4.1 Description of the experimental system*

An experimental system based on the description provided in section 2 was built and a control and sensing system was prepared. Figure 10 displays the system including the rotor. The stator is built from an aluminum annulus connected to three Langevin piezo actuators with horns, where each actuator is driven by a voltage amplifier. The rotor, which is an ABS disk, is connected to a rotational bearing at it midpoint, confining it to a concentric rotational movement. The rotor is carved to reduce its weight but still fully cover the annulus and provide a firm connection to the bearing. A magnetic encoder (contactless) is placed above the rotation axle to measure the angular position of the rotor. The system is controlled by an FPGA digital board, handling the generation of input signals to the actuators' power driver, processing the encoder's reading and applying the control law.

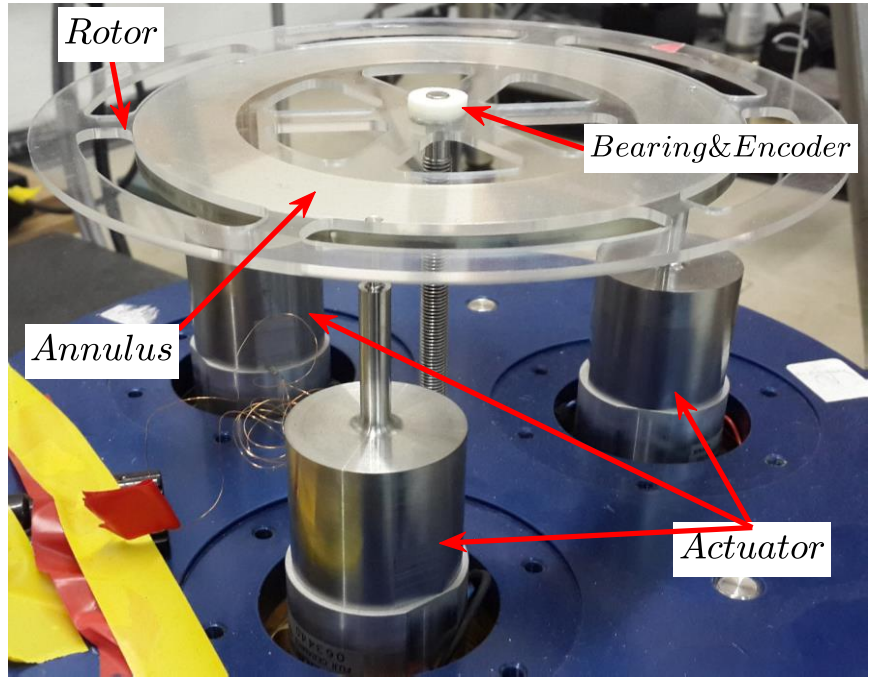

Figure 10. Experimental System.

*4.2 Open-loop experiments*

4.2.1 Traveling wave generation

Initially, it was desired to verify that the design could indeed generate the expected vibration patterns. Thus, the annulus was excited at the natural frequency of the desired mode with eleven nodal diameters $(k=11)$. The vibrations were measured using a scanning laser vibrometer (Polytec$^{TM}$, OFV-303) along its circumference at several radii. Figure 11 shows a measured spatial vibration of the annulus.

The ability to impose a desired wave (with the desired value of *SWR*/*TWR*) is a key feature when applying a feedback control. The latter was tested by measuring the spatial response of the annulus (with the scanning laser) for different values of phases for the actuators and calculate the values of *SWR* and *TWR* by applying the ellipse method [16]. Examples of the measured and fitted ellipses for three different actuation phases are presented in Figure 12. Indeed, the well-fitted ellipse model (See [23]) suggests that a single wavelength mode is excited without additional contaminations.

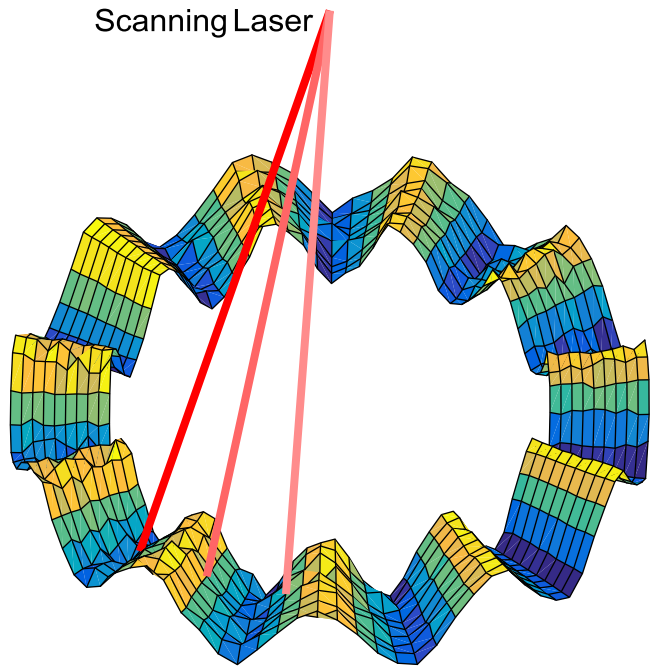

Figure 11. Measured vibrations of the annulus scanned by the Laser Vibrometer.

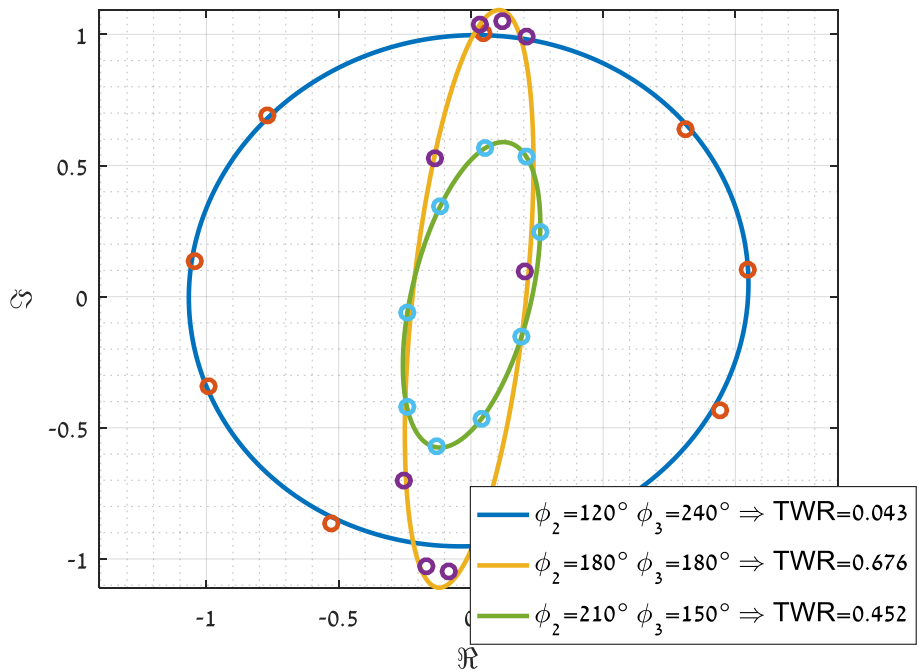

Figure 12. Measured real vs imaginary parts of the complex amplitude along the ring. The proportions of the ellipse are related to the different *TWR* values.

By exploring the entire range of *u* from -1 to 1, the *SWR/TWR* are obtained as a function of the phase ratio between the actuators (*u*). The results are presented in Figure 13 and compared to the simulated values from Figure 7(a). The measured *TWR/SWR* are very close to the simulated curves. Two differences appear in the boundary values. (i) The measured *TWR* is slightly asymmetric with respect to the selected curve of *u* vs. the actuators phase shift (from Eq.(27)). (ii) The *TWR* values do not achieve values of either $TWR = 1$ (pure traveling wave) or $TWR = 0$ (pure standing wave). The source of both differences is the variation of amplitudes between the actuators as presented in the analysis of section 3.2, and Figure 8. A calibration procedure in which the phase parameter *u* to phase shift is adjusted to the measured *TWR* overcomes these differences and improve the system's behavior. The calibration procedure searches the actual location on the *TWR* map where $TWR = 0$ (Figure 5 is an example map). The search is carried out on-line by tuning the amplitudes and phases of two actuators relative to the 3rd one. After the exact locations of the traveling waves are found, it is possible to define the *u* function line.

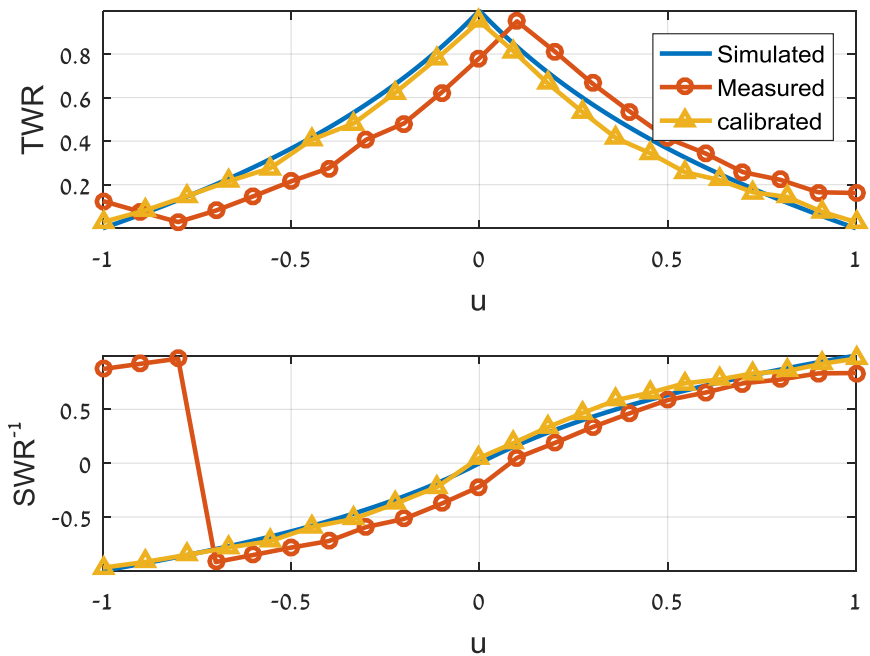

Figure 13. Comparison of simulated, measured and calibrated *TWR* (top) or $SWR^{-1}$ (bottom) vs. the phase shift (*u*).

The calibration process produces a modification to Eq.(27) in a way that remaps *u* to the path on the map that is closest to the simulated *TWR*. In this experiment, it turns out that the calibrated Eq.(27) is:

$$\begin{aligned}\varphi_2(u) &= (3/10 \cdot u + 31/30)\pi \\ \varphi_3(u) &= (-3/10 \cdot u + 31/30)\pi.\end{aligned} \qquad (28)$$

Figure 13 also presents the calibrated function of *TWR/SWR*. It is shown that after calibration, the obtained functions of *TWR/SWR* are close to the desired function allowing smooth and continues regulation of the waves.

### 4.2.2 Dynamic modeling verification

It is important to verify the relations between the applied traveling wave (via *u*) to the rotating torque that is being developed by the acoustic layer. At steady state, the inertial forces $(J\ddot{\theta})$ are canceled (at Eq.(23)) and the rotating torque balances the dissipation forces. This balance leads to a constant steady-state velocity that is directly related to the rotating torque magnitude. Thus, Eq. degenerates into:

$$c_0 + c_1\dot{\theta} + c_2 \operatorname{sign}(\dot{\theta})\dot{\theta}^2 = M(u). \tag{29}$$

By exciting propagating waves with different *TWR* (by setting different values of $u$) and letting the rotor reach steady-state rotation speed, the relation between the *TWR* and the applied torque can be evaluated. Figure 14 presents the results of such an experiment. The obtained rotating velocity is slightly asymmetric, with respect to the applied wave (via $u$), as observed earlier. In addition, it is shown that for small values of $u$ there is no rotation. This is the effect of static friction in the bearing.

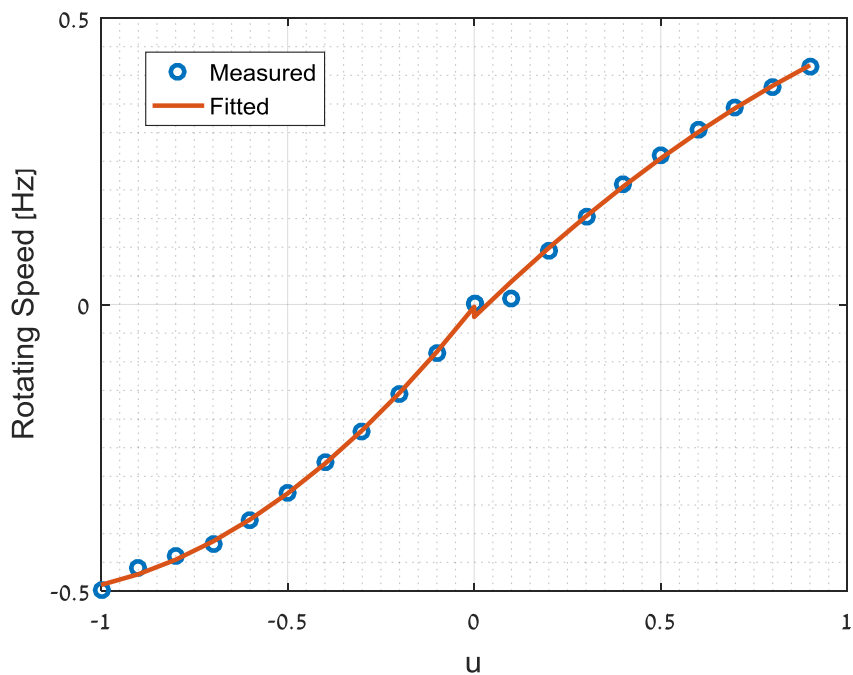

Figure 14. Steady state rotating speed for various values of *u*.

Applying the relation between $u$ and the $SWR^{-1}$, obtained in Figure 13, the relation between the traveling waves and the steady state velocity can be calculated. Since *SWR* is directly related to the generated rotating moment, according to Eq.(29), it is possible to fit a parabolic function, so $SWR^{-1} = \alpha_0 + \alpha_1\dot{\theta} + \alpha_2 \operatorname{sign}(\dot{\theta})\dot{\theta}^2$. The free term $\alpha_0$ is related to the coulomb damping of the bearing, the linear term $\alpha_1\dot{\theta}$ is related to the viscous damping, originated in the air, and the quadratic term $\alpha_2\dot{\theta}^2$ is related to the drag forces. The experimental measurements and the fitted curves of steady-state velocity vs. $SWR^{-1}$ are shown in Figure 15.

By decomposing the linear and quadratic terms and preserving only the quadratic term, only the drag forces related effect remains. At steady state, drag is balanced by the propelling torque to yield the steady rotation speed. It appears that the measurements favorably compare with the simulations results presented in section 2.3 (Figure 4). The same information is also shown in Figure 15 by the dashed line. Comparing this result with Figure 4, the same dynamic behavior is revealed. The differences occur since both the levitation gap and the frequency are not constant, nor are they controlled during the experiment.

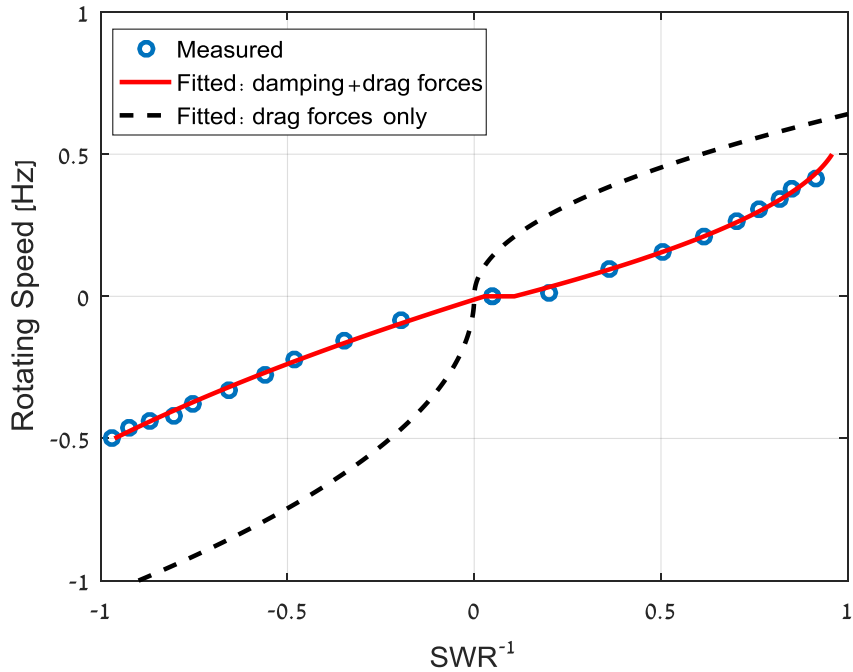

Figure 15. Steady state rotating velocity vs. $SWR^{-1}$. The dashed line is without the linear damping.

## *4.3  Motor Transfer function*

In order to design a controller, a model of the motor plant (rotating moment to position) is required. The transfer function was measured by exciting the motor with a harmonic input, generating a harmonic rotating moment applied to the rotor:

$$u = U_0 \sin(\omega t). \tag{30}$$

Measuring the motor's angular oscillations $\theta(t)$ under periodic change of the relative phase, *u,* for a range of frequencies, allows calculating the magnitude and phase lag for each frequency. A 4$^{th}$ order transfer function model was fitted to the measurements. The model is composed of a double integrator (the dynamics of the rotor inertia), a natural frequency at 3.7Hz and, a complex zero. At the vicinity of the natural frequency the response magnitude is attenuated by 40dB, adding some phase delay. The measured frequency response and the fitted model are shown in Figure 16.

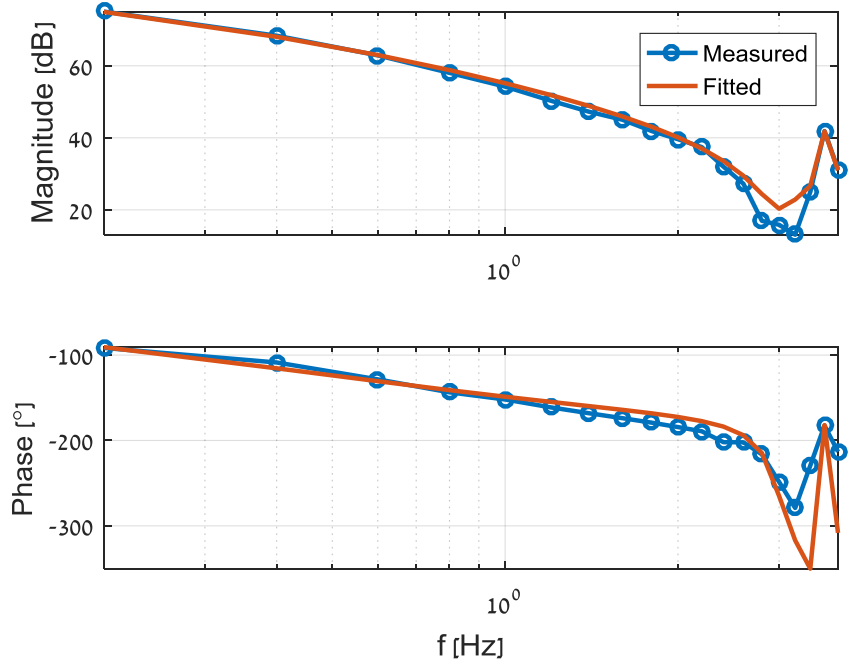

Figure 16. Open loop frequency response.

### *4.4 Closed loop control*

A position control scheme can be utilized by feeding back the encoder reading (the rotor angular position) to the control law that is computed on the FPGA digital card. A PID controller was designed to stabilize the system and hold the reference position. Figure 17 presents the closed-loop response for a step position command. The purpose of this experiment is to prove the concept of closed-loop control with the acoustic levitation traveling wave actuation. The controller successfully stabilized the rotor position with some overshoot. The steady state error is due to the fact that low rotating moments (where $TWR \approx 1$) do not overcome the bearing friction while the rotor is nearly stationary. Careful design of a controller can improve the closed-loop performances significantly beyond what is shown here.

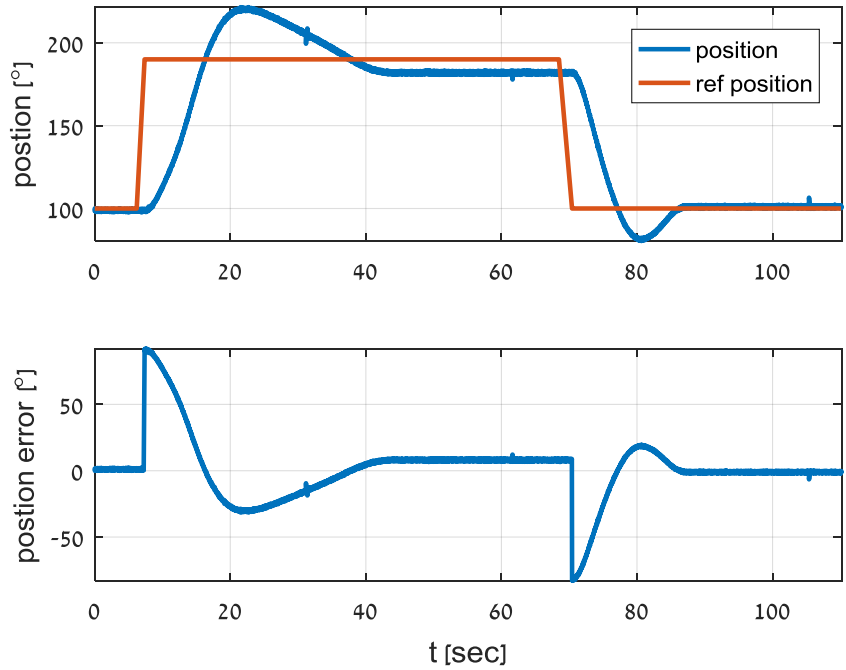

Figure 17. The response of the Closed-loop controlled system to a step position command.

## 5 Conclusions

A contactless ultrasonic motor based on acoustic levitation and traveling wave control was presented. The rotation control scheme generates the necessary pressure field that gives rise to an acoustic levitation force and the rotational moment. A novel actuation configuration reducing over actuated system into a single control parameter was developed. This configuration is designed to maintain the axisymmetric structure of the system, while enabling the efficient transfer of electrical to mechanical power to the system. By controlling the amplitude and phase of each actuator individually, different proportions of traveling to standing flexural waves are obtained allowing regulating the rotational moment. The analysis shows that additional calibration is required to overcome structural imperfections and to effectively generate traveling waves. The rotational moment was shown to be related to the quality of the traveling wave to standing waves ratio. Thus, the ability to control the purity of the traveling wave allows to control the angular position of the rotor provided a position feedback. The described system has verified the theory and simulations were proved by a series of experiments that was presented.

## Acknowledgements

This research was funded by the Ministry of Science, Technology, and Space.